\documentclass[apj]{emulateapj}

\slugcomment{Astrophysical Journal - in press}

\shorttitle{POLARIZATION REVERBERATION MAPPING}
\shortauthors{GASKELL ET AL.}

\begin{document}

\title{DISCOVERY OF POLARIZATION REVERBERATION IN NGC 4151}

\author{C. MARTIN GASKELL\altaffilmark{1,2,3}, RENE W. GOOSMANN\altaffilmark{4},
NELLY I. MERKULOVA\altaffilmark{5,6},\\ NIKOLAY M.
SHAKHOVSKOY\altaffilmark{5}, AND MASATOSHI SHOJI\altaffilmark{1,2}}

\altaffiltext{1}{Department of Physics \& Astronomy, University of
Nebraska, Lincoln, NE 68588-0111. }

\altaffiltext{2}{Astronomy Department, University
of Texas, Austin, TX 78712-0259. \email{mshoji@astro.as.utexas.edu}}

\altaffiltext{3}{Present address: Centro de Astrof\'isica de Valpara\'iso y Departamento de F\'isica y Astronom\'ia,  Universidad de
Valpara\'iso, Av. Gran Breta\~na 1111, Valpara\'iso, Chile. \email{martin.gaskell@uv.cl}}

\altaffiltext{4}{Observatoire astronomique de Strasbourg, 11 rue de l'Universit\'e, F-67000 Strasbourg, France.
\email{rene.goosmann@astro.unistra.fr}}

\altaffiltext{5}{Crimean Astrophysical Observatory, Nauchny, Crimea,
98409 Ukraine}

\altaffiltext{6}{Deceased: 2004 December 12}

\begin{abstract}

Observations of the optical polarization of NGC~4151 in 1997--2003
show variations of an order of magnitude in the polarized flux while
the polarization position angle remains constant.  The amplitude of
variability of the polarized flux is comparable to the amplitude of
variability of the total $U$-band flux, except that the polarized flux
follows the total flux with a lag of $8 \pm 3$ days. The time lag
and the constancy of the position angle strongly favor a scattering
origin for the variable polarization rather than a non-thermal
synchrotron origin.  The orientation of the position angle of the
polarized flux (parallel to the radio axis) and the size of the lag
imply that the polarization arises from electron scattering in a
flattened region within the low-ionization component of the
broad-line-region.  Polarization from dust scattering in the
equatorial torus is ruled out as the source of the lag in polarized
flux because it would produce a larger lag and, unless the
half-opening angle of the torus is $>$53\degr, the polarization would be
perpendicular to the radio axis. We note a long-term change in the
percentage of polarization at similar total flux levels and this could be
due either to changing non-axisymmetry in the optical continuum
emission, or a change in the number of scatterers on a
timescale of years.

\end{abstract}

\keywords{galaxies:active --- galaxies:quasars:general ---
polarization --- galaxies:individual:NGC 4151 --- galaxies:Seyferts}


\section{INTRODUCTION}

\citet{shklovskii53} pointed out that synchrotron radiation should
be highly polarized, and the detection in 1953 of polarization of
the Crab Nebula demonstrated the synchrotron nature of its continuum
radiation \citep{dombrovskii58}.  It was therefore natural to search
for polarization of the optical emission of Seyfert galaxies
\citep{dibai_shakhovskoy66}.  Optical polarization of NGC~4151 was
discovered by \citet{babadzhanyants_hagen-thorn69} who also reported
that the polarization was variable.  The polarization was studied in
more detail by \citet{kruszewski71} who found that the polarization
in the $V$ band varied by a factor of four over a nine-month period.
The detection of variable polarization was taken as evidence of a
synchrotron origin of the continuum.

The discovery of polarization of emission lines \citep{angel76} in
NGC~1068 showed that the polarization of type-2 AGNs was due to
scattering and not synchrotron emission.  The origin of the
optical polarization in type-1 AGNs, however, has remained ambiguous.
\citet{stockman79} discovered that the optical polarization of
radio-loud AGNs was parallel to the radio jet axis. They considered
this to be either the result of intrinsic polarization of optical
synchrotron emission, or the result of scattering from a flattened
distribution of scatters. Subsequently \citet{antonucci82,
antonucci83} showed that, in general, polarization tends to be
parallel to the radio axis in type-1 AGNs and perpendicular in type-2
AGNs.  Since the polarization of broad lines is roughly similar to
the polarization of the continuum \citep{goodrich_miller94} this
showed that this polarization was due to scattering.  It was shown by
\citet{smith04} and \citet{goosmann+gaskell07} that type-1
polarization can be explained by a flattened equatorial scattering region.  Nonetheless
because of the variability of optical polarization it has often been
assumed, by analogy with blazars (where there is no doubt that the
optical polarization variations have a non-thermal origin), that
synchrotron radiation also contributes to optical polarization in
non-blazar AGNs (e.g., Giannuzzo \& Salvati 1993).  In this paper we
attempt to resolve the question of the nature of the optical
polarization by investigating the relationship between the
variability of the polarized flux and the total flux.

\section{OBSERVATIONS}

NGC~4151 was observed through a 15-arcsecond diameter aperture with
the UBVRI Double Image Chopping Photometer -- Polarimeter
\citep{piirola88} on the 1.25-m AZT-11 telescope of the Crimean
Astrophysical Observatory. The polarimeter gives simultaneous
measurement of the linear polarization in the standard Johnson bands
by using dichroic filters. The resulting passbands have effective
wavelengths of 3600, 4400, 5400, 6900, and 8300 \AA ~which are
close to the standard Johnson $UBRVI$ system. Observations of
NGC~4151 were made on 58 nights between May 1997 and May 2003. About
30 separate measurements were made per night.    Errors in the flux
and polarized flux were estimated by comparing measurements taken
within $\pm 5$ nights of each other.

\section{RESULTS}

\citet{merkulova06} have already discussed the variability of the total flux and polarized flux in the five passbands.  As is normal when there is substantial host galaxy contamination because of the use of a large photometric aperture, the highest amplitude total flux variability is in the $U$-band.  This is shown in Fig.~5 of \citet{merkulova02}.  In NGC~4151, as in other AGNs, the $B$-band and $U$-band fluxes vary essentially simultaneously.  \citet{crenshaw96} found that the UV continuum variations in NGC~4151 are simultaneous to
within $\sim \pm 0.15$ days across the UV and \citet{edelson96} get an upper limit of $\sim 1$ days to the lag of the $\lambda$5100 continuum with respect to the continuum at $\lambda$1275 (see Fig.~10 of \citealt{lyuty05}).  Wavelength-dependent lags are seen at longer wavelengths in the $R$ and $I$ bands \citep{lyuty05,sergeev05} but \citep{gaskell07} argues that these are due to contamination by emission from hot dust. We do not expect a significant lag of the $B$ band relative to the $U$ band and there is no observational evidence for one.  \citet{oknyanskij+03} find the $V$ band to be lagging the $B$ band by $+0.2 \pm 0.4$ days, and \citet{sergeev05} find it to be lagging by $-0.07 \pm 0.22$ days.  The last two limits are important because the former overlaps the first third of our polarimetric monitoring and the later the last third.  We can thus be confident that there is no significant delay between the $U$ and $B$ bands and that variations in $U$ are an excellent indicator of variations in $B$.

The polarization was variable in all five pass bands and it too might be expected to be highest towards shorter wavelengths, but \citet{merkulova06} find that both the polarization and polarization variability of NGC~4151 peak in the $B$ band and show a significant drop in the $U$ band. This is found in other objects.  \citet{kishimoto04} find from their spectropolarimetry that the polarized flux spectrum of AGNs commonly peaks in the region of the $B$ band and drops off in the $U$ band. They attribute this to a Balmer absorption edge in the accretion-disk emission.  Detailed modelling shows that the spectrophotometry of the Balmer edge region is well fit with a Doppler broadened Balmer absorption edge \citep{gaskell09b}.

Because the polarization variability of NGC~4151 is at a maximum in the $B$ band, and because the signal-to-noise ratio is also highest in the $B$ band (because of the amplitude of the variability and the high quantum efficiency of the detector in the $B$-band), we only consider the variability of the $B$-band polarization here. We show the variability of the $B$-band polarized flux and the variability of the total flux (taken from the $U$-band) in Fig.~1.

\begin{figure}
\includegraphics[width=0.48\textwidth]{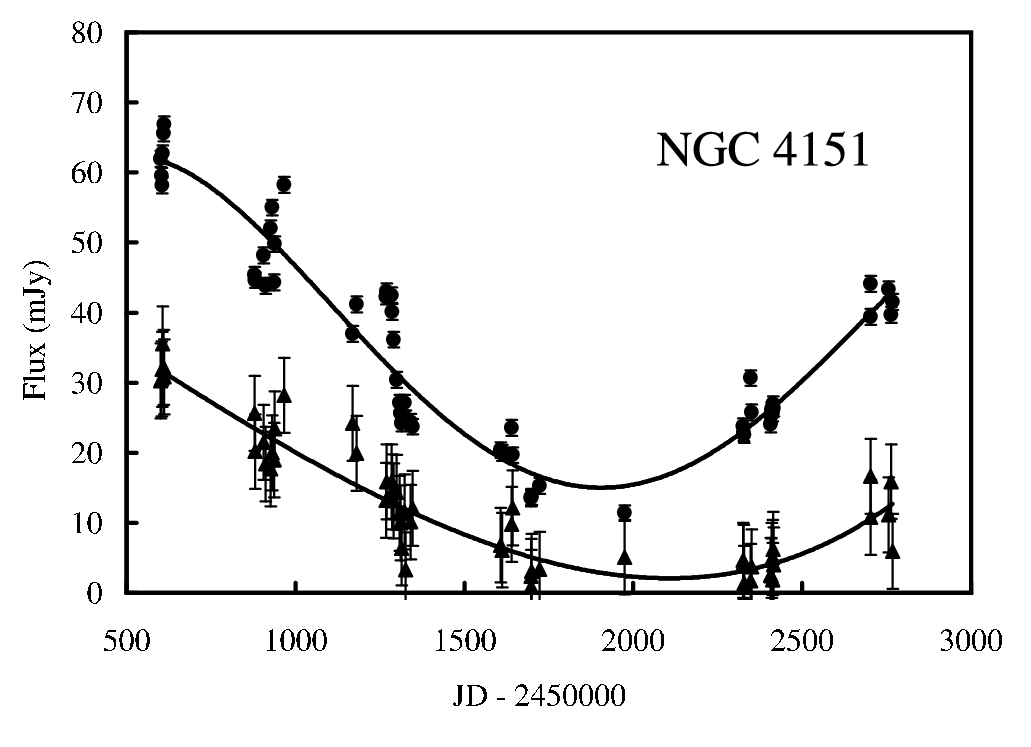}
\figcaption{Variations in total $U$-band flux (top curve) and $B$-band polarized flux (lower curve).  The polarized flux has been multiplied by a factor of 30 for plotting convenience.  The polarization position angles in both wave-bands are consistent with each other and remained nearly constant at $92\degr \pm 1.6\degr$ during the whole observational campaign.  The two curves are fourth-order polynomial fits through the data.}\label{fig1}
\end{figure}

As can be seen from Fig.\@ 1, during the period 1997--2003 the $U$-band flux showed long-term variability of at least a factor of four, and the $B$-band polarized flux showed long-term variability of a factor of ten. \citet{merkulova06} find that the position angle (PA) of the variable polarization remained nearly constant at 92\degr $\pm$ 1\fdg 6 which is consistent with earlier measurements \citep{antonucci83,martel98} and with being parallel to the radio axis \citep{antonucci83}.

\section{ANALYSIS}

\subsection{Long-Term Variability}

During the period 1997 to 2003 NGC~4151 declined from the end of a
high state and passed through a low minimum level of activity
\citep{lyuty06}. There is some evidence for a slow long-term change
in the degree of polarization over the seven-year period. This is
most obvious if we compare years where the mean $U$-band continuum
level was approximately the same. As can be seen from Fig.\@ 1, the
polarized flux around MJD 2400 is more than a factor of two lower
than that around MJD 1300 when the total fluxes are comparable.
Similarly the polarized flux is also lower around MJD 2750 compared
with MJD 1200. This change in the percentage polarization after the
photometric minimum around MJD 2000 suggests a possible reduction on
a dynamical timescale in the number of scatterers after the low
state. Another possible explanation, strongly favored by other observations
(see \citealt{gaskell10} and \citealt{gaskell11} for extensive discussion), is that the
continuum variability could be non-axisymmetric and the anisotropy
could be varying from year to year.  Unfortunately we did not have
good polarimetric coverage in the year of the photometric minimum.

\subsection{Short-Term Variability}

In order to search for polarization reverberation on short
timescales we have removed the long-term trends by subtracting out
fourth-order polynomials (shown in Fig.\@ 1) from the total flux and
polarized flux time series.  The two time series were
cross-correlated using the interpolation method of
\citet{gaskell_sparke86}.  Details of the method are given in
\citet{gaskell_peterson87}.  We show the resulting cross-correlation
function (CCF) for the entire data set in Fig.\@ 2.
\begin{figure}
\includegraphics[width=0.48\textwidth]{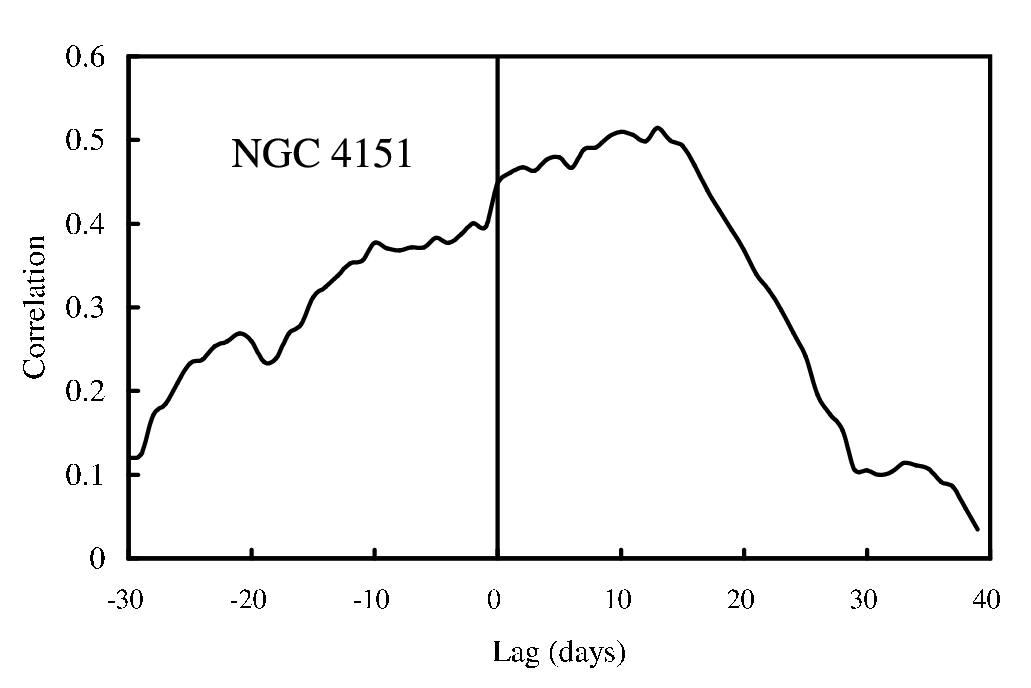}
\figcaption{Cross-correlation function for the polarized flux and
total $B$-band flux for the entire six-year data set.  A positive
lag corresponds to the polarized flux following the total
flux.}\label{fig2}
\end{figure}

The centroid of the CCF calculated from the first moments above 0.7
times the peak correlation gives a lag of 8 days. In Table 1 we also
give the considerably less certain lags from the centroids of the
CCFs for the individual observing seasons.  The median of the lags
for the individual years is consistent with the lag from the whole
of the seven observing seasons.  The differences in lags in Table 1
are not statistically significant, but if the variability is
non-axisymmetric \citep{gaskell10,gaskell11} future better observations could
show real variations in the polarization lags vary from year to year.

As is well known \citep{gaskell_peterson87}, the distribution of
errors in the peaks in cross-correlation functions is non-Gaussian
with a tail extending to high errors.  This can be seen in the
simulations of \citet{maoz_netzer89}, \citet{white_peterson94}, and
\citet{peterson98}.  We estimated the error in the lag of the
polarized flux in three different ways. First we used the formula of
\citet{gaskell_peterson87}.  This gave an error of $\pm 3.5$ d for
the lag derived from the whole data set. We also performed Monte
Carlo simulations where we added noise to the observations equal to
the observational errors.  This gave an error in the lag of $\pm
2.5$d.  Finally, where possible we calculated the lags for the
individual years as described above.  They are again the first moment lags calculated
above 0.7 of the peak in the CCF.  These lags are shown in Table 1
with the approximate errors from the \citet{gaskell_peterson87}
formula. As can be seen, even though the errors are substantially
larger because of the small number of observations, the median lag
for the individual years (8.5 d) is close to the lag we obtained
from all years.  The median of the absolute values of the
differences of the individual lags for each year from the lag for
all years is 1.5 d. These three separate error estimates suggest
that the error in the lag for the whole sample is of the order of
$\pm 3$ d.

\begin{deluxetable}{ccc}
\tablewidth{0pt} \tablecaption{Centroids of Lags for Individual
Years} \tablehead{ \colhead{Year} & \colhead{Lag (days)} &}
\startdata

1997 & 7 & $\pm4$ \\
1998 & 43 & $\pm6$ \\
1999 & 7 & $\pm7$ \\
2000 & 10 & $\pm5$ \\
2001 & --- & \\
2002 & 14 & $\pm8$ \\
2003 & 7 & $\pm5$ \\
All years & 8 & \\
\enddata
\vspace{0.3cm}
\end{deluxetable}

\section{DISCUSSION}

A $\sim 8$ d lag of the polarized flux has no natural explanation if
the polarization has a synchrotron origin.  However, a lag can
naturally be explained as a light-travel-time delay of scattered
radiation \citep{giannuzzo_salvati93}. As noted above, scattering
off both electrons and dust has long been considered as a cause of
polarization in AGNs. Electron scattering could take place in clouds
along the axis of symmetry or in a flattened equatorial distribution
of electrons \citep{smith04,smith05}.  Dust scattering in the torus will
also be responsible for polarization. To investigate which process
dominates in producing the polarized flux we modeled polarization
reverberation in NGC~4151 with the Monte Carlo radiative transfer
code $STOKES$ \citep{goosmann+gaskell07,goosmann07}. In order to
model polarization variability we modified $STOKES$ to time-tag each
photon going through the program.

\subsection{Polarization from the Dusty Torus}

For the period 1990--1998 \citet{oknyanskij99} find a delay of $35 \pm 8$ days for the $K$-band relative to the $U$ band.  This includes the peak of activity at the start of our polarimetric monitoring. For the low state during our monitoring \citet{minezaki04} find a similar delay of $48 \pm 2.5$ days, and by the end of our observations \citet{Swain2003} constrained the radial extension of the {\it K}-band emitting region to 0.04~pc, which also corresponds to a time-lag of $\sim 48$ days. \citet{kishimoto2009} found an identical {\it K}-band size in 2009. Note that \citet{Swain2003} believed that the {\it K}-band emission comes from the outer accretion disk while \citet{kishimoto2007,kishimoto2009} attribute it to the inner boundary of the torus. The difference in both interpretations may lie in the assumed dust sublimation radius. If one assumes a grain-size distribution favoring larger grains, the dust can survive at a relatively close distance of 0.04~pc from the central source. Indications that dust grains in AGN tend to be larger than in standard Milky Way dust are found from examining quasar reddening curves \citep{gaskell2004}.

\citet{oknyanskij93} did determine a $K$-band lag during the low state of NGC~4151 in the 1970s of $18 \pm 6$ days which is marginally consistent with the polarization reverberation delay we measure, but as \citet{oknyanskij06} point out, their observations and those of \citet{minezaki04} show that, as would be expected, the high state of NGC~4151 at the start of our polarimetric monitoring destroyed the dust close to the black hole, and the $K$-band observations show that the dust had still not reappeared at smaller radii within several years of the high state.

Overall, there is good observational evidence that the inner radius of the torus did not change significantly during our monitoring that finished in May 2003.  \citet{koshida2009} report that the inner boundary of the torus moved in after 2003 but when \cite{pott2010} and \citet{kishimoto2011} took new interferometry data, the radius of the {\it K-}band emission region was again at $\sim 0.04$~pc from the center.  The $K$-band emission comes from the hottest dust, and lags given by the cross correlation method are biased towards material at the smallest radii \citep{gaskell_sparke86} so the $K$-band lag is giving the inner radius of the dust torus.  From the \citet{oknyanskij99} and \citet{minezaki04} measurements we can conclude the {\em inner} radius of the dusty torus during our polarimetric monitoring was $\sim 40$ light days.  This strongly rules out dust scattering being the cause of the polarization reverberation we detect.

We modeled dust scattering off a variety of optically-thick torus
geometries. We considered cylindrical tori and tori with elliptical
cross sections.  In all cases we set the inner radii to be 40 lt-d
as indicated by the IR reverberation mapping.  Various opening and
viewing angles were modeled.  The resulting first moment delays in
the polarized flux were $\sim 43$ days (i.e., slightly greater than
the inner radius), and varied only by 10--15\% with changing
geometries and viewing angles.  These modeled lags are clearly
inconsistent with the $\sim 8$ d lag we find from the observations.
So long as the torus was not too thin (i.e., the half-opening angle
is not $>$ 53\degr) the polarization of the dust model averaged over the lag was
perpendicular to the radio axis for all the type-1 viewing angles,
which is inconsistent with the observed position angle.  We thus
believe that scattering from the torus is not the cause of the lag
in the polarized flux. For more extensive discussion of the effect
of the geometry and viewing angle on the degree and direction of
polarization see \citet{goosmann+gaskell07}.

\subsection{Polarization from a Flattened Electron Scattering Region}

Scattering off a polar distribution of electrons
\citep{giannuzzo_salvati93} also produces the wrong polarization
angle, but a flattened electron distribution produces polarization
parallel to the radio axis (see discussion in \citealt{goosmann+gaskell07}).
We therefore used $STOKES$ to model the polarization lag
from cylindrical electron-scattering disks.  We found that for disks
which were optically thick to electron scattering the lag was equal
to the inner radius.  Thus if the electron scattering region is
optically thick in NGC~4151, it has an inner radius of $\sim 8$
lt-d.

The geometry of the scattering region is similar to that of a BLR
(see review by \citealt{gaskell09a}).  Furthermore, the lag we find is
comparable to the size of the BLR of NGC~4151.
\cite{gaskell_sparke86} obtained a size of $5 \pm 2$ lt-d for the
radius of the C~IV emitting region of the BLR in NGC~4151 during
1978-1980. \cite{metzroth06} obtained radii of $3.4 \pm 1.3$d for
1988 and $3.3 \pm 0.9$d for 1991. They also obtained identical radii
for He\,II. \cite{gaskell_sparke86} estimated the size of the region
emitting H$\beta$ and H$\gamma$ to be $\sim 6$ lt-d during
1980-1981. More recent observations by \citet{bentz06} give $6.6 \pm
1$ lt-d for 2005.  Note again that the responsivity-weighted radii
given by reverberation mapping are biased towards the inner radii of
the emitting region.

While our observed polarization lag is in good agreement with the observed radii of the C IV and H$\beta$ emission in NGC 4151, the observed inner radius for a given line reflects the radial variation of ionization and there is almost certainly BLR gas inside that radius. If the gas density of the electron scattering region in our $STOKES$ models is somewhat lower than a canonical BLR density (i.e., $10^8$ cm$^{-3}$ rather than $10^{10}$ cm$^{-3}$), then the model is no longer optically thick, the mean free path to electron scattering becomes significant, and we find that the lag is greater than the inner radius of our hypothetical disk. For example, a disk with a density of $10^8$ cm$^{-3}$ would give a mean free path of $\sim 6$ lt-d. It is therefore possible to reproduce the observed polarization lag without having to have an artificial hole in the middle of the distribution of electrons.

Interestingly, the radius we obtain for the electron-scattering disk (i.e., towards the outer edge of the BLR) is in good agreement with the location deduced quite independently by \citet{smith05} from modeling the change in polarization with velocity across the profiles of broad emission lines. Nevertheless, it is more intuitive to assume a continuous accretion flow from the torus down to the supermassive black hole.

\section{Conclusions}

The polarized flux of NGC~4151 appears to lag the unpolarized flux by $\sim 8$ d which is comparable to the light-crossing time of the broad-line region.  We interpret this as the result of the extra light-travel time of the scattered photons. Dust in the torus is ruled out as the source of polarization by the angle of polarization and because the observed lag is too short. Both the direction of polarization and the lag can naturally be explained instead by electron scattering in a flattened region of similar size to the low-ionization BLR. We have also detected a possible change in the percentage polarization of NGC~4151 at similar flux levels on a timescale of years, and we attribute this to either long-term changes in the number of scattering electrons, or variable asymmetric emission in the disk.

It is clearly important to confirm these results with better data.  We
believe that further polarimetric reverberation mapping of AGNs (especially
spectropolarimetric reverberation mapping) should be a powerful
technique for distinguishing between models of the structure of the
inner regions of AGNs.

\acknowledgments

We grateful to Makoto Kishimoto for helpful comments. This research has been supported by the US National Science Foundation through grants AST 03-07912 and AST 08-03883, the University of Nebraska UCARE program, the
GEMINI-CONICYT Fund of Chile through project N{\degr}32070017, and the French {\it GdR} PCHE. Polarimetric observations at the Crimean Astrophysical Observatory were made possible in part by
Award No.\@UP1-2549-CR-03 of the U.S. Civilian Research \& Development Foundation (CRDF).

\end{document}